**Diagnosing Coronal Heating Processes with Spectrally Resolved Soft X-ray Measurements**

Amir Caspi[1], Albert Y. Shih[2], Harry P. Warren[3], Marek Stęślicki[4], & Janusz Sylwester[4]
[1]Southwest Research Institute, Boulder;  [2]NASA Goddard Space Flight Center;
[3]Naval Research Laboratory;  [4]PAS Space Research Centre

Decades of astrophysical observations have convincingly shown that soft X-ray (SXR; ~0.1–10 keV) emission provides unique diagnostics for the high temperature plasmas observed in solar flares and active regions. **SXR observations critical for constraining models of energy release in these phenomena can be provided using instruments that have already been flown on sounding rockets and CubeSats.** These instruments have relatively low cost and high TRL, and would complement a wide range of mission concepts.

The solar corona, at quiescent (non-flaring) temperatures of ~1–10 MK, is ≳100× hotter than the underlying chromosphere and photosphere. This "coronal heating problem" remains one of the fundamental unanswered questions in solar physics (see, e.g., [1]). Magnetohydrodynamic simulations and observations of convective flows (e.g., [2] and references therein) suggest that the Sun's complex magnetic field is an efficient conduit for energy transport from the solar interior and subsequent storage in the corona, but the mechanism for releasing that energy to heat the corona remains unknown.

Models based on impulsive dissipation of magnetic complexity through magnetic reconnection ("nanoflares," e.g., [3]) suggest that coronal plasma should be routinely heated to flare-like, ~5–10 MK temperatures, but with relatively low density (e.g., [4,5]). In contrast, models based on dissipation of Alfvén waves predict relatively narrow, cooler coronal temperature distributions (e.g., [6,7]). Observations showing hot emission from the active Sun (e.g., [8–10]; and Fig. 1) and much cooler emission from the quiet Sun (e.g., [11]) appear to support nanoflare heating for active regions but a different mechanism – such as small-scale flux cancellation (e.g., [12]) – for the quiet network. However, the difficulty of measuring weak, high-temperature emission – particularly using extreme ultraviolet (EUV) observations – has led to inconsistent results and multiple, conflicting interpretations (e.g., [13–16]).

Solar flares produce copious high-temperature plasma at temperatures up to ~30–50 MK (e.g., [17,18]). While easily measurable, the physical mechanism that drives this heating nonetheless remains poorly understood. It is commonly accepted that much of the flare thermal plasma results from chromospheric material "evaporating" into the corona as it is heated by collisions from non-thermal, downward-accelerated electrons [19,20]. However, numerical simulations of this process (e.g., [21,22]) have difficulty reproducing the "super-hot," >30 MK plasma observed in intense flares. Indeed, a growing body of evidence (e.g., [17,18,23–27]) suggests that a significant fraction of the thermal plasma – especially in the hot tail of the temperature distribution – is heated *in situ*, directly in the corona. This hottest plasma, particularly the super-hot component, exhibits fast dynamics during the impulsive phase and can precede the onset of hard X-ray (HXR) emission from the flare-accelerated electrons [17,19], suggesting its sensitivity to the details of the energy release process (e.g., [18,25]). How-ever, the exact mechanism for *in situ* heating, and its relationship to non-thermal particles, remains strenuously debated (e.g., [20,22,28]).

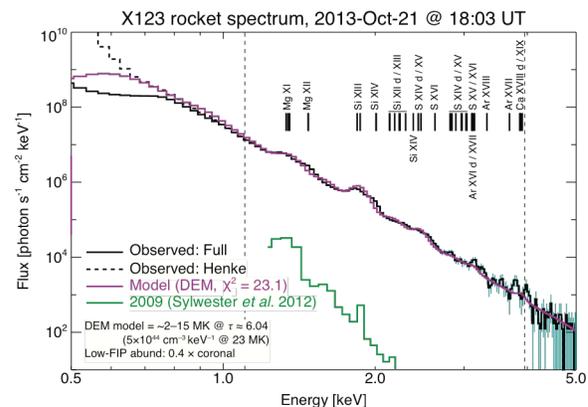

Fig. 1. SXR spectra support nanoflare heating in active regions, showing high-temperature emission from an active Sun (*black*) versus only low temperatures in the quiet Sun (*green*). Prominent spectral lines show variations from normal coronal abundances [10].

Measurements of elemental abundances in hot coronal plasma provide crucial additional information on how mass flows within, and into, the corona in response to heating. It is well established that the composition of the solar atmosphere varies from photosphere to corona (see [29] for a review), with variations organized by first-ionization potential (FIP) whereby low-FIP abundances tend to be enhanced in the corona relative to the photosphere (the "FIP bias").

Abundance measurements of hot plasma thus test models of plasma origin, but studies so far have yielded mixed results, complicated by the aforementioned observational difficulties (including possible non-equilibrium ionization) and possibly by the different temperature sensitivities of the lines studied. For example, recent sounding rocket SXR observations revealed a near-photospheric composition for high-temperature quiescent active region plasma [10] (Fig. 1), while prior EUV studies have generally shown coronal FIP biases (e.g., [15,16]). Moreover, spatially resolved SXR spectra of two simultaneous active regions using a novel imaging spectrograph (discussed below) suggest a significantly higher O-to-Fe ratio in one region versus the other, despite roughly similar temperature distributions [30] (Fig. 2). In flares, some studies have shown a photospheric abundance for Fe but an enrichment for (lower-FIP) Ca (e.g., [31,32]), while others show variations from flare to flare (e.g., [33,34]). An inverse FIP effect (Ar enhanced relative to Ca) has also been detected in sunspots [35,36]. This variability may suggest that the fractionation threshold could depend on the details of the heating mechanism and the properties of the ambient magnetic field, although these seemingly disparate results could also potentially be reconciled by evolution of the FIP bias over active region lifetime (e.g., [37,38]). Systematic studies are clearly needed to make progress on this question, but abundance measurements from many flares and active regions over long periods of time have been very difficult to make with previous instrumentation.

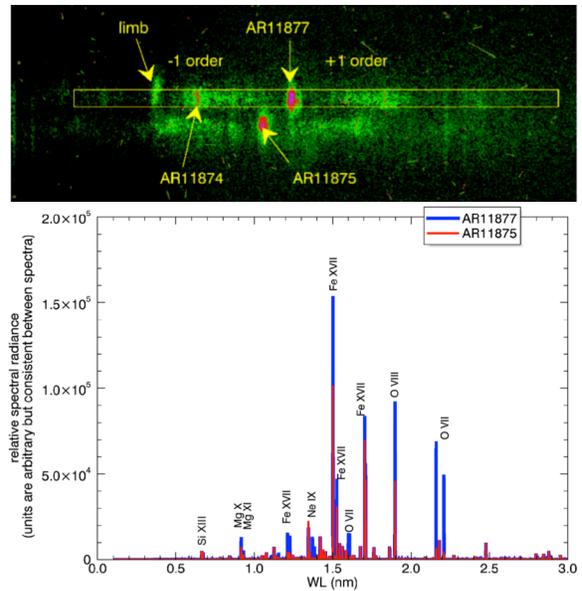

**Fig. 2.** Variations in active region abundances could indicate differences in heating mechanisms or region age. [*Top*] MOXSI prototype observation of multiple active regions, with clear spectral dispersion. [*Bottom*] Derived spectral models for the two brightest observed regions show markedly different line ratios [30].

**All of these questions are ideally addressed via high-resolution SXR spectroscopy**, both spatially integrated and spatially resolved. SXR emission is particularly sensitive to mid- and high-temperature plasma, ~2–50 MK, and includes strong emission lines from both low- and high-FIP elements (Fig. 1, 2) across this temperature range, from O VIII (~2 MK) to Fe XXVI (≳25 MK). Spectrally resolved observations thus provide an ideal diagnostic of both quiescent and flaring coronal temperature distributions and composition [39].

**Significant progress can be made with only modest spectral, spatial, and temporal resolutions and signal-to-noise (SNR) requirements.** $E/\Delta E$ or $\lambda/\Delta\lambda$ of only ≳25 and SNR ≳10 is sufficient to resolve prominent spectral line clusters across a range of temperatures and FIPs (e.g., O VIII, Mg XI, Si XIII, Ca XIX, Fe XVII & XXV) and to achieve ~20% accuracy on simultaneous temperature distribution and relative abundance determinations. A cadence of ≲30 s during

flares resolves thermal dynamics during the impulsive phase, and ≲1 hr during quiescence samples even short-timescale variability. Spatial resolution of just ~50″ resolves individual active regions, while ~10″ can resolve individual flare features.

**These observational constraints can be met with relatively low-cost and low-resource instruments that require no technology development.** Spatially integrated spectroscopy using silicon drift detectors (SDDs) has already been proven for solar observation on sounding rockets [10] and on the *MinXSS* CubeSat [40,41]. These SDDs use thermoelectric coolers within a vacuum housing with a beryllium entrance window to provide ~0.15 keV FWHM spectral resolution over ~0.5–30 keV. They can be easily complemented by proven cadmium-telluride (CdTe) detectors to extend the energy range for spectroscopy well into the HXRs (up to ~100 keV).

Spatially resolved spectroscopy can be achieved from a novel slitless diffractive imaging spectrograph concept, the Multi-Order X-ray Spectral Imager (MOXSI). This technologically simple design provides spectro-spatial images over ~0.2–10 keV by dispersing the SXR emission using a transmission grating and overlaying the dispersed spectra on the (wavelength-integrated) $0^{th}$-order image. The concept has been proven with a sounding rocket prototype [30] (Fig. 2); it is similar to the *Chandra* HETG instrument [42], but operates on spatially resolved solar features (e.g., active regions and flares) rather than point-like stars, reminiscent of the Skylab SO-82A "overlappograph" [43]. To provide a $0^{th}$-order baseline for data analysis, and for additional spectral information, MOXSI includes additional apertures without a transmission grating but with thin filters, offset from the primary aperture. This yields SXR filtergrams similar to those made by the *Hinode* XRT [44], enabling temperature diagnostic capabilities even for the quiet Sun.

The simulated MOXSI image in Fig. 3, based on a CubeSat-scale design, illustrates the expected observation. The tremendous intensity of solar SXR flux, even during quiescence, enables direct imaging up to ~10 keV (~1.2 Å) using just a pinhole aperture and short focal distances. MOXSI on *CubIXSS* [45], a proposed NASA CubeSat, achieves ~25″ and ~0.25 Å FWHM resolutions with a *Chandra* HETG flight-spare grating and ~25 cm focal length. Ready enhancements include improved sensitivity and resolution via longer focal lengths (1.5 m yields ~7″ and ~0.1 Å) and/or replicated focusing optics at low cost and negligible risk.

**These high-TRL, low-resource instruments enable unprecedented SXR observations for studying solar flares and active regions, and complement a wide range of mission concepts.**

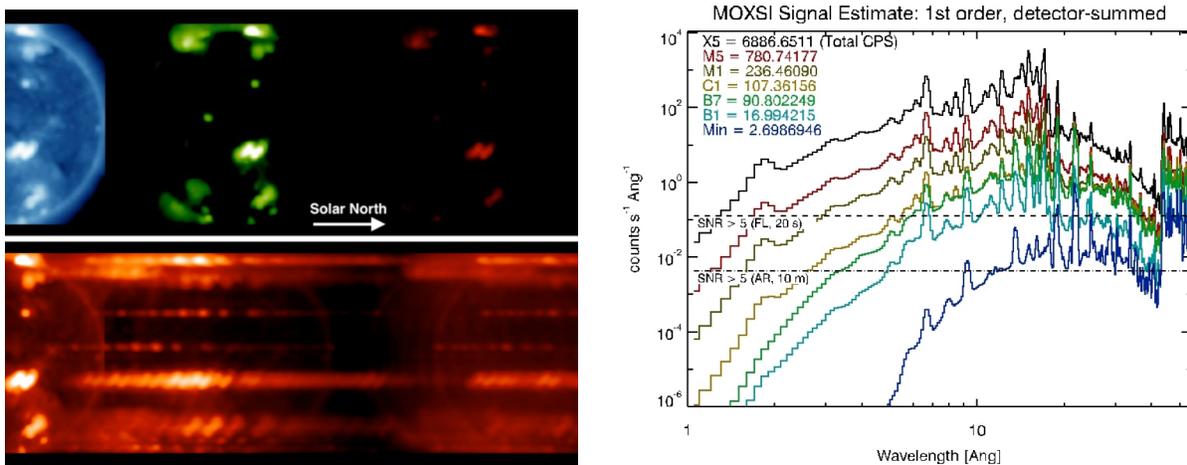

**Fig. 3. MOXSI can make unprecedented measurements of thermal plasma in flares and active regions.** [*Left*] Representative observation simulated from *Hinode*/XRT data forward-folded through the MOXSI response, including filtergrams (top) and dispersed image (bottom). Only half (+$1^{st}$ order) of the detector is shown for brevity. [*Right*] Estimated detector-integrated dispersed $1^{st}$-order (spectral) signal for various solar conditions. Examples shown are for CubeSat-scale instrument; a meter-class instrument would further improve sensitivity and spectro-spatial resolution.